\documentclass[reqno,12pt]{hdsr}

\graphicspath{{figs/}}

\begin{document}

\newgeometry{bottom=1.5in}


\begin{center}

\textbf{\LARGE {Paschen curve for a PCB trace in low vacuum}}
\normalsize

   \thispagestyle{empty}
   \vspace*{.2in}

  \begin{tabular}{cc}
    N. Atanov, S. Bini, S.Ceravolo, G. Corradi, R. Donghia 
   \\[0.25ex]
   \end{tabular}
  
  \emails{
    Corresponding author: N. Atanov, atanov@jinr.ru 
    }
  \vspace*{0.4in}
\end{center}

\textbf{Abstract}\\
For one PCB trace region, where bare high voltage trace goes near a GND pad, we estimated an electrical breakdown voltage in a low vacuum using simplified geometrical model under some assumptions. Experimental measurement of Paschen curve for the PCB board considered was proceed in pressure range from 2x10\textsuperscript{-3} mbar to 25 mbar.

\vspace*{0.15in}
\hspace{10pt}
  \small	
  \textbf{\textit{Keywords: }} {Paschen law, PCB, vacuum}

\section{Introduction}
\label{sec1}
For microelectronics devices with high voltage traces that work in a vacuum environment it is important to know real dependece of breakdown voltage on pressure to avoid malfunction. Paschen law is well known equation for breakdown voltage behavior when a pressure and a distance change. It’s common mathematical expression [1] is written under uniform field assumption for two parallel conductive plates. There are some recent works where a non-uniform electrical filed is considered for several special conductor configurations and also for PCB traces in a vacuum with pressure up to 10\textsuperscript{-1} mbar [2]. There are also reports about anomaly of Paschen curve behavior for uniform field, very low distances (~ 10 um and closer) and low vacuums [3,4]. 
Here we introduce investigation of Paschen effect for non-uniform fields that occurs for one common PCB trace configuration with distances ~100 um in low vacuums up to 10\textsuperscript{-4} Torr. In Section 2 of this paper we provide a simplified theoretical estimation of minimum breakdown voltage using Townsend criteria. In Section 3 experimental setup to measure breakdown voltage dependence on pressure is described and in Section 4 results of experimental investigation for the PCB trace in vacuum camera are presented.

\begin{figure}[H]
    \centering
    \includegraphics[scale=0.35]{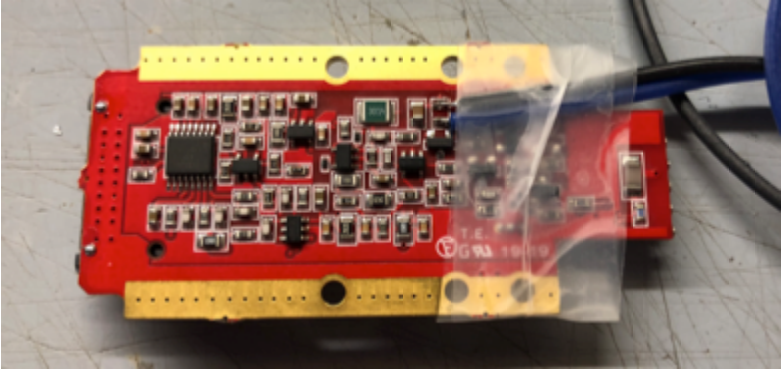} \\
    \caption{Double sided multilayer analog PCB board that was used for tests. HV is driven with cable.}
    \label{fig:my_label}
\end{figure}

\section{Theoretical estimation}

For tests we used multilayer double-sided printed circuit board for analog nodes (Fig.1). We perform analysis of places where breakdown can occur, and for further consideration we selected the most critical section of this PCB layout (Fig.2). Here are bare HV and GND traces on the surface of board under connector that have air layer between them. The minimum distance between traces is 300 um, the sizes of traces are 500 um in length for HV and 1500 um for GND, the trace diameters are 200 um for HV and 300 um for GND. 

\begin{figure}[H]
    \centering
    \includegraphics[scale=0.35]{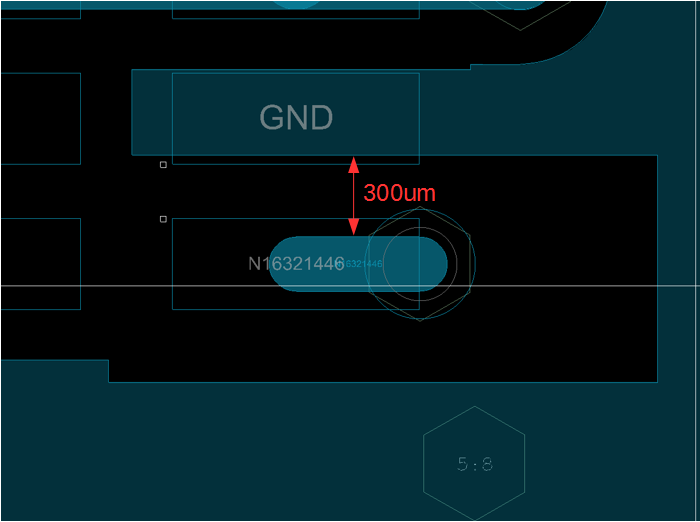}
    \includegraphics[scale=0.35]{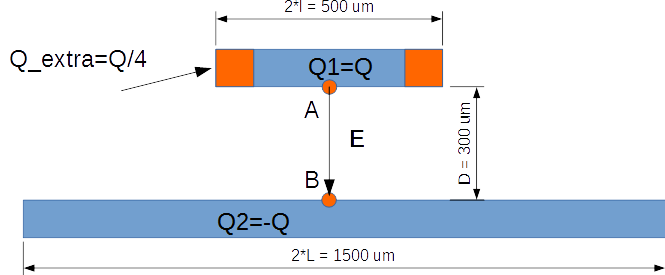} \\
    \caption{The most critical path is considered: distance between HV trace and ground trace is 300 um.}
    \label{fig:fig_2}
\end{figure}

In general, one need solve Poisson equation for a space region to get charge and field distribution and it is connected with a complex calculation. But to get qualitative estimation how non-uniform field impacts on a breakdown let us consider a simplified model of traces geometry (Fig.2 right) and estimate a charge, a field and a breakdown voltage for this configuration. We calculate the field for trace AB as the highest field strength area. 

Primary assumptions on this model are:

\hspace*{1cm}-	traces are assumed to be long 1-dimensional threads with linear charge density λ;

\hspace*{1cm}-	system is in an equilibrium and we have Q1=-Q2 due to charge conservation law.

Moreover, for further we assume also that [1]:\\
\hspace*{1cm}- cathode has enough free electrons that can be accelerated and start ionization;

\hspace*{1cm}-	further free electron growing is only due to ionization;

\hspace*{1cm}-	each ionized atom creates only one free electron;

\hspace*{1cm}-	free electrons on the surface of cathode are created with ion’s colliding.

In fact, for l/D ~1 one can show that there is an extra charge on the edges, that depend on l/D [5], and for our sizes total 
Qtotal = Q + Qextra= 1.5Q. In addition, to conserve a charge one should increase charge density in the GND trace: ${\lambda}_{2} = \lambda + \Delta\lambda = (Q+Qextra)/2L = 1.5 \lambda$.
To calculate field first let us calculate a basic field from linear charge distribution and then add edge extra charge field as a field from point charges.

For linear charge distribution due to electric filed superposition:
\textbf{E = E1 + E2 + Eextra}, 		where E1, E2 are field strength for upper and lower trace, Eextra is the field from point charges on edge.

Therefore, for every PCB trace on Fig.2(right) we have: 

\begin{equation}
    \mathbf{E_{1,2}} = \int^l_{-l}\mathbf{dE} = \int^l_{-l}\mathbf{dE\bot} + \int^l_{-l}\mathbf{dE\|} = \int^l_{-l}\mathbf{dE\bot}
\end{equation}

Here  $\int^l_{-l}\mathbf{dE\|} = 0$ due to symmetry of system considered (Fig.3).
And for a given distance from trace \textit{y} we have:

\begin{equation}
    E_1 = \frac{\lambda}{4\pi{\epsilon}_0}\int^l_{-l}\frac{\cos{\alpha}dx}{r^2} = \frac{\lambda}{2\pi\epsilon_0}\frac{l}{y\sqrt{l^2+y^2}}
\end{equation} 
Similarly, one can find the $E_2$ and extra charge filed. So the total field module $E$ will be:

\begin{equation}
\begin{split}
    E = E_1 + &E_2 + E_extra =\\ 
    &=\frac{\lambda l}{2\pi\epsilon_0}\left({\frac{1}{y\sqrt{l^2+y^2}} + \frac{1.5}{(D-y)\sqrt{L^2+(D-y)^2}} + \frac{y}{2\sqrt{(y^2+a^2)^3}}}\right)
\end{split}
\end{equation}

\begin{figure}[H]
    \centering
    \includegraphics[scale=0.35]{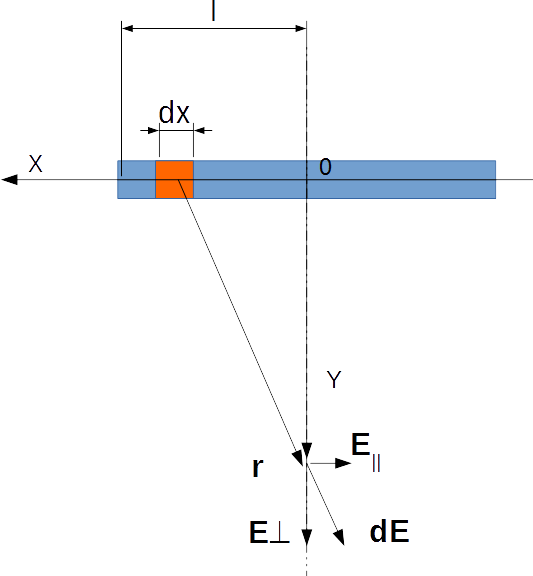} \\
    \caption{Calculation of a field strength for the trace.}
    \label{fig:fig_3}
\end{figure}

Next, it is necessary to find a charge. We know the value of voltage applied across two traces, so we can do it, using the equation:

\begin{equation}
\begin{split}
U = \Delta\phi &= \int^D_0 E_ydy = \\ 
        &= \lambda\int^D_0\frac{1}{2\pi\epsilon_0}\left(\frac{1}{y\sqrt{l^2+y^2}} + \frac{1.5}{(D-y)\sqrt{L^2+(D-y)^2}} + \frac{y}{2\sqrt{(y^2+a^2)^3}}\right)dy
\end{split}
\end{equation}
And for calculated $\lambda$ the resulting plot for electrical field strength module along y coordinate is shown on Fig.4.

\begin{figure}[H]
    \centering
    \includegraphics[scale=0.35]{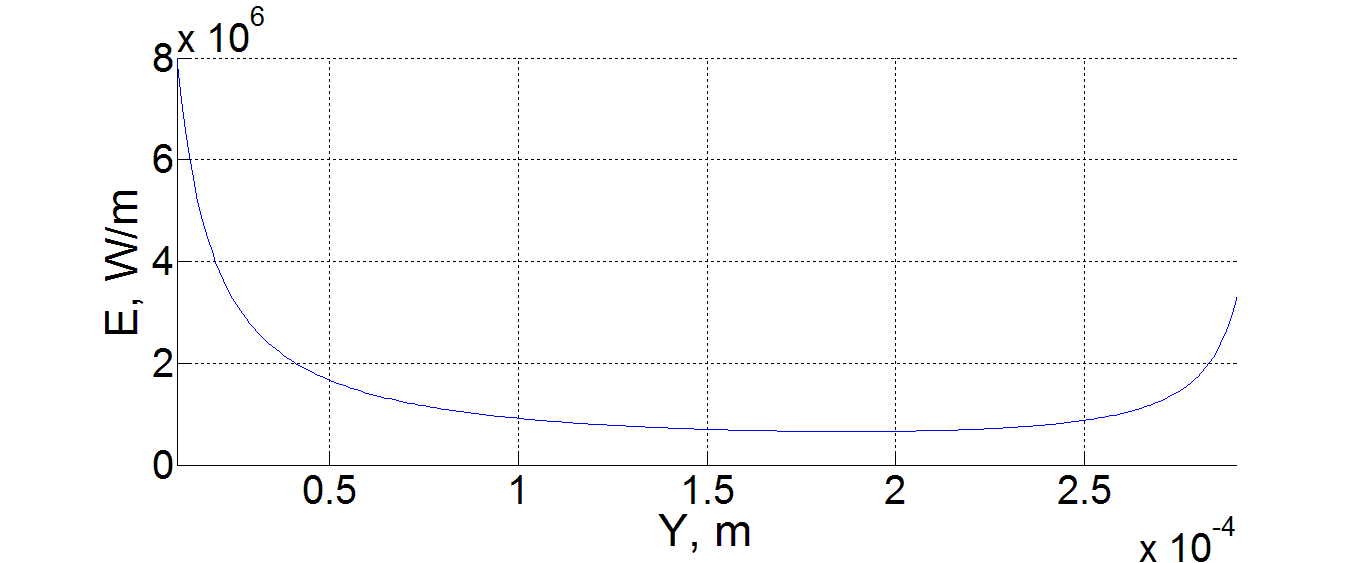} \\
    \caption{Calculated field strength for AB trace of the PCB considered.}
    \label{fig:fig_4}
\end{figure}

Now, it is possible to compare our configuration of PCB traces with well known two plates configuration with uniform electrical field that is used to build classical Paschen curve equation.As we know for extended space region a breakdown in gas can be described with the Townsend criteria [1]:

\begin{equation}
    \int^D_0\alpha(y)dy \geq \ln{\left(1 + \frac{1}{\gamma_{se}}\right)}
\end{equation}
where
  $\alpha(y)=Ape^{-Bp/E(y)}$, and A and B are constants, that depends on gas.
  
  For uniform field it becomes 
  
\begin{equation}
ApDe^{-\frac{BpD}{V}} \geq \ln{\left(1 + \frac{1}{\gamma_{se}}\right)}
\end{equation}

Here, ${\gamma}_se$ is the secondary electron emission coefficient, and it is constant in some pressure range. 
For air and pD=5.7 Torr*mm (minimum pD value for Paschen curve), if p or D are not very low, we know that A=112.5 kPa/cm and B=2737.5 V/(kPa*cm) and Vbreakdown = 327V. So for whole space region considered with non-uniform field the breakdown criteria can be written as\\
$\\ \int^D_0\alpha(y)dy \geq 254.59\\$

And for our PCB trace configuration at the same condition (pD = 5.7 Torr*mm, where D=300um) on can easy find with eq. (3,4,5), that this condition is right when applied voltage\\
Vbreakdown $>$ 630 V.\\
\\
I.e. in our case, non-uniform field from the PCB traces leads to breakdown voltage increasing.


\section{Experimental measurements. Hardware}
Next step is to measure breakdown voltage in a real board as a function of pressure. The main sources of breakdown are PCB trace on the top of boards and connector pins, high voltage was applied with cable to connector and traces. To provide interface outside the vacuum camera these wires are soldered to short coaxial cable with LEMO connector. 

\begin{figure}[H]
    \centering
    \includegraphics[scale=0.55]{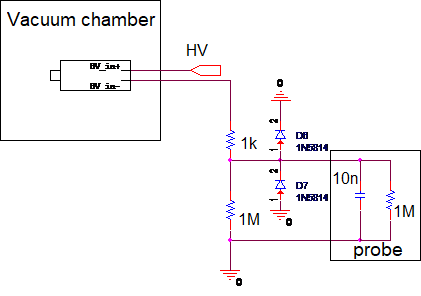}
    \includegraphics[scale=0.35]{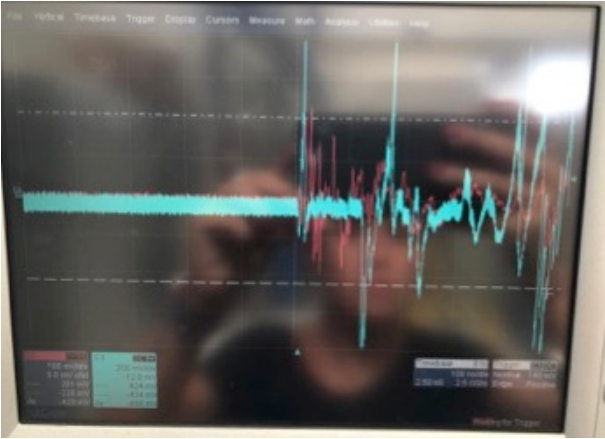} \\
    \caption{Breakdown detection scheme: at the breakdown moment corona occurs and HVin- pin goes to HV providing a pulse that can be observed with a scope probe.}
    \label{fig:fig_5}
\end{figure}

The HV is supplied with programmable VIM HV module that have short circuit protection on board, indicating breakdown and current higher than threshold level.

Another tool to observe corona during breakdown was used. It is a special probe (Fig.5), serially connected to ground and \verb+HV_In-+ pin of the PCB. Normally, \verb+HV_In-+ have GND potential. When breakdown occurs, the circuit begins to conduct, current flows and \verb+HV_In-+ pin goes to high voltage. Protection diodes limit a voltage and one can safely observe a voltage pulse with oscilloscope probe, setting the trigger on this event.

\section{Measurements results}
After the circuit has been installed and checked, the chamber was closed and the pressure inside goes down with aid of vacuum pump. On several point it stopped and breakdown voltage measured by increasing the HV and observing the breakdown. We used two methods described above: with a trigger on the scope and observing the short circuit protection signal on HV power supply block. Typical probe response for a breakdown is shown on Fig. 5 (right). The results are provided in table on the Fig.6 and on the plot on the Fig. 7. 

One can see minimum breakdown voltage V=405 V at the point 5 mBar. This value for preamp is higher, than known value for air breakdown minimum with two flat electrodes Vair = 327 V at pD= 5.7 Torr*mm. For pressure point p=$10^{-3}$ – $10^{-2}$ mBar the breakdown voltage is higher than 1600 V. And it seems to go to saturation, that is in good coincidence with known result for two flat electrodes and low pressures and distances of 100 um order [4].

\begin{figure}[H]
    \centering
    \includegraphics[scale=0.35]{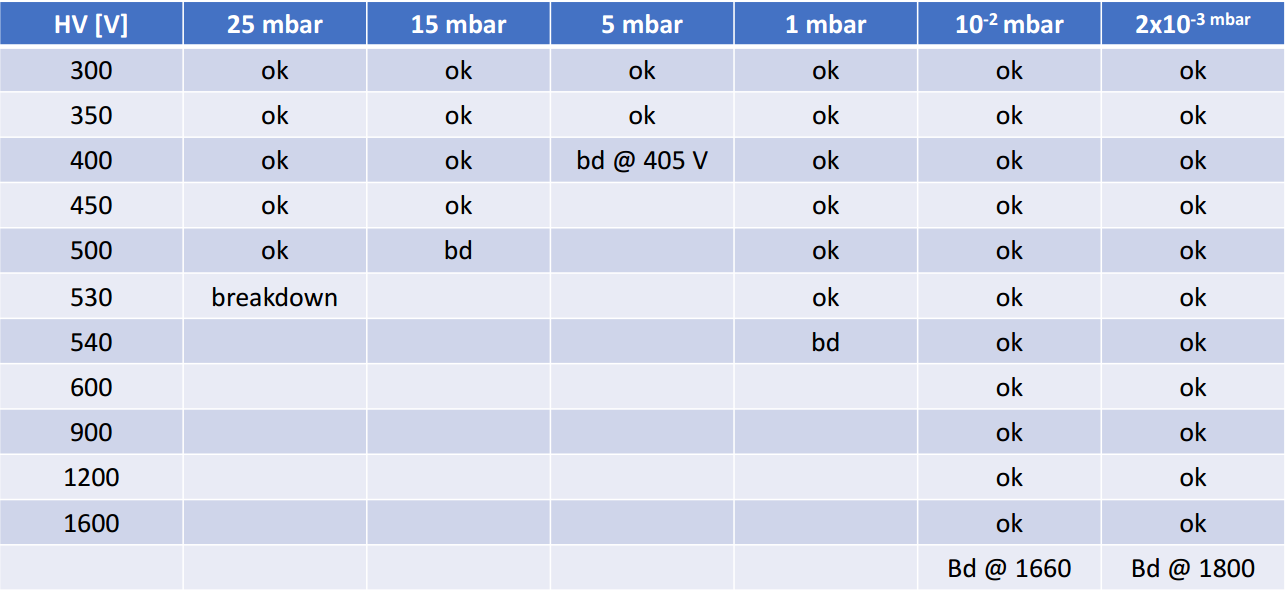} \\
    \caption{Dependence of breakdown voltage on the pressure in vacuum chamber.}
    \label{fig:tab_1}
\end{figure}

\begin{figure}[H]
    \centering
    \includegraphics[scale=0.35]{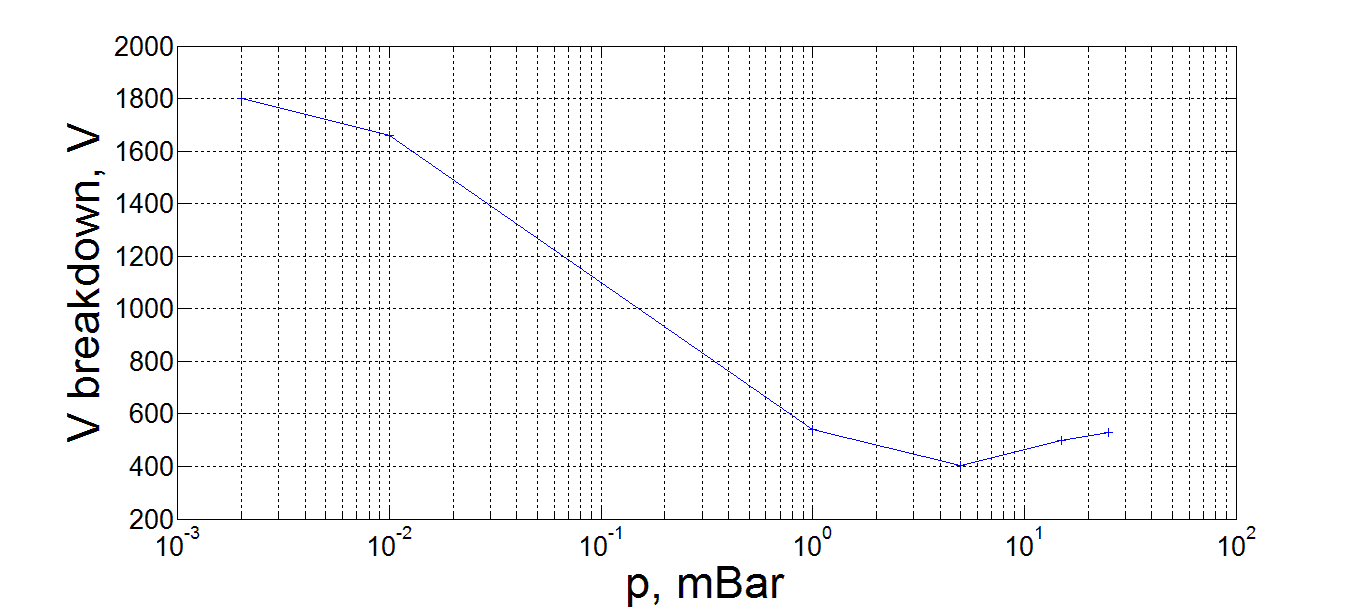} \\
    \caption{Breakdown detection scheme: at the breakdown moment corona occurs and HVin- pin goes to HV providing a pulse that can be observed with a scope probe.}
    \label{fig:fig_6}
\end{figure}

\section{Conclusion}
We have estimated field strength, charge distribution and breakdown voltage using simplified geometry of PCB traces. The qualitative estimation gives us that a breakdown voltage grows as a field became more non-uniform due to finite sizes of conductors.

We performed experimental measurement of breakdown voltage dependence on a pressure in vacuum camera in range 2x$10^{-3}$ mbar to 25 mbar. Measurement suggests that breakdown voltage is higher for finite length conductors

One can see saturation in low vacuums less than $10^{-2}$ mBar that is not predicted by Paschen law equation.


\end{document}